# Evolution of the Lyα forest: a consistent picture

**Jan P. Mücket**[1], **Patrick Petitjean**[2,3], **Ronald E. Kates**[1], **and Rüdiger Riediger**[1]

[1] Astrophysikalisches Institut Potsdam, An der Sternwarte 16, D-14482 Potsdam, Germany
[2] Institut d'Astrophysique de Paris - CNRS, 98bis Boulevard Arago, F-75014 Paris, France
[3] UA CNRS 173- DAEC, Observatoire de Paris-Meudon, F-92195 Meudon Principal Cedex, France

**Abstract.** We study the evolution with redshift of the number density of Lyα clouds in a CDM model using numerical simulations including photo-ionization and cooling of the baryonic component. The photo-ionizing flux is consistently taken as proportional to the rate at which material cools below 5000 K in the simulation. The number density of lines with $N$(HI) $> 10^{14}$ cm$^{-2}$ given by the simulation can be approximated as $dn/dz \sim 16.8(1+z)^{0.8}+0.687(1+z)^{2.5}$ which fits well the data at any redshift. The number of weaker lines is reproduced within a factor of two and does not decrease with time as fast as the number of strong lines. For $N$(HI) $> 10^{13}$ cm$^{-2}$, we find $dn/dz \propto (1+z)^{0.7}$.

At $z = 0$, the two-point correlation function for Lymanα clouds with $N$(HI) $> 10^{13}$ and $> 10^{14}$ cm$^{-2}$ has a slope ~ -1.75 comparable to the slope of the two point correlation function for the galaxy distribution. The signal is higher for stronger lines for separations smaller than 1 Mpc. Part of the Lymanα forest is indeed associated with galaxies however the overall picture is that Lyα absorption lines originate in the warm gas that traces the potential wells of filamentary structures defined by the dark matter.

## 1. Introduction

Observation of QSO absorption line systems is the most sensitive way to probe baryonic material in the whole redshift range $0 < z < 5$. Among absorption line systems, Lyα systems without detected associated metal lines are the most numerous (about 150 per unit redshift at $z \sim 2$). They have been extensively studied, since they may trace the evolution of the intergalactic medium and the UV background. Indeed the clouds could be an intergalactic population possibly of primordial gas evolving independently of galaxies under the direct influence of the intergalactic medium, the ionizing background, and the gravitational potential of dark matter. They are observed over a large redshift range, and they thus probe the evolution of the baryonic material and the structures formed by dark matter.

The correct picture for the origin of the Lyα forest will most likely involve a mixture of all the ideas invoked up to now, and probably the Lyα forest arises through clouds in very different environments. Gas weakly enriched in heavy elements could be correlated with galaxies due to a variety of mechanisms: First, when galaxies form, the energy released could eject a substantial amount of gas at high velocity and thus to a comparatively large distance from the center (Lake 1988). This gas could be weakly enriched in heavy elements. Second, after formation, an initial starburst phase would inevitably lead to supernovae, creating a strong temperature gradient, ejecting metal-enriched gas, and thus again enhancing the medium surrounding the galaxy with metals. While large galaxies would be expected to undergo numerous starburst phases, resulting in roughly solar abundances, dwarf galaxies might be expected to have only one significant starburst phase, resulting in a weak metal enhancement in surrounding clouds. Third, during the process of merging, which seems to be a common event in the course of galaxy evolution, metal enriched gas could be ejected very far from the center (Morris & van den Bergh 1994), but still close enough to be associated with galaxies and thus show some clustering.

However, primordial gas should be anticorrelated with star forming regions. The subpopulation of intergalactic clouds with primordial abundances should evolve independently as pressure confined gas (Ikeuchi & Ostriker 1986), or gravitationally confined gas in the potential wells of mini-halos (Ikeuchi 1986, Rees 1986), such as those predicted in a cold dark matter (CDM) scenario. The model of gravitational collapse on subgalactic scales (Bond et al. 1988) has received some support from recent hydrodynamic simulations (Cen et al. 1994) suggesting that Lyα clouds could result from pancake structures (Zel'dovich 1970) in which the photoionized gas is maintained by a balance between gravitation and ram pressure due to infalling gas.

It is therefore important to clarify the link between absorption line systems and galaxies. This can be done by studying

*Send offprint requests to*: J. Mücket

the content of the gas in metals, the clustering properties of the systems, and the properties of the galaxies close to the line of sight to the quasar.

It is generally assumed that the metallicity is very small (Chaffe et al. 1986), but this assumption is not firmly established. In fact, the presence of CIV in Ly$\alpha$ systems has been claimed by Lu (1991), who finds evidence for the presence of the CIV$\lambda$1548 line in a composite spectrum obtained by shifting and stacking 172 Ly$\alpha$ systems of equivalent widths $0.4 < W_{\rm r} < 0.8$ Å. A similar study but using higher resolution data leads to an upper limit of one-hundredth of solar abundances (Tytler & Fan 1994). Very recently weak CIV lines associated with Ly$\alpha$ lines of HI column density $10^{14.5}$ cm$^{-2}$ have been detected (Tytler 1995, Cowie et al. 1995) leading to [C/H] > -2.5 for those clouds. However no firm conclusion can be drawn from these detections. It shows that at least part of the Ly$\alpha$ forest contains metal but the crucial observation would be to put strong limits on abundances at lower column density where the bulk of the Ly$\alpha$ forest is to be found (Petitjean 1995a). Indeed the break in the $N$(HI) column density distribution at $N$(HI) $\sim 10^{15}$ cm$^{-2}$ has been interpreted as evidence for two populations of Ly$\alpha$ systems (Petitjean et al. 1993), the weakest ($N$(HI) $< 10^{14}$ cm$^{-2}$) and more numerous could be low metallicity intergalactic clouds and the strongest ($N$(HI) $> 10^{15}$ cm$^{-2}$) would be related to galactic haloes. In such a picture it is not surprising to find metals in the latter population. C IV and O VI doublets have been searched for in low-redshift Ly$\alpha$ systems with no detection down to $W_{\rm r} < 8$ and 10 mÅ respectively (Petitjean 1995b).

Galaxies are strongly clustered and if the absorbing gas is associated with galaxies, absorption line systems are expected to cluster on scales smaller than 1000 km s$^{-1}$. The issue has drawn much attention but is still unclear. At high redshift ($z > 2$), the distribution is practically uniform on scales larger than 300 km s$^{-1}$. On smaller scales, there has been some indication for clustering of lines with $W_{\rm r} > 0.36$ Å (Webb 1987, Ostriker et al. 1988, Crotts 1989, Barcons & Webb 1991). Since the bulk of the Ly$\alpha$ forest has smaller equivalent width however, high spectral resolution data should be used. Rauch et al. (1992) found no evidence for clustering using a sample of 295 lines with $N$(HI) $> 10^{13.75}$ cm$^{-2}$. Using similar data, Chernomordik (1995) claimed a detection of clustering on scales 50-150 km s$^{-1}$ for lines with $N$(HI) $> 10^{14}$ cm$^{-2}$. The fact that the properties may be column density dependent seems to be confirmed by analysis of recent new high resolution data (Cristiani et al. 1995). Great care has to be exercized however since the clustering signal can be diluted if the dimensions of the clouds are not negligeable compared to their mean separation (Bajtlik 1995).

Sizes can be derived from detections of lines coincident in redshift in spectra of QSOs with small projected separations on the sky. The Ly$\alpha$ forest appears strikingly similar in the two images of the lensed quasar Q2345+007 separated by 7.3" (Foltz et al. 1984, Smette et al. 1992) indicative of large transversal dimensions. Only loose constraints are obtained with that object on the radius of the absorbers, $R > 2h^{-1}$ kpc. Stronger limits have been obtained from observation of brighter objects. HE1104-1805A & B are separated by 3" and coincidences imply $R > 50h^{-1}$ kpc for $W_{\rm r} > 0.085$ Å (Smette et al. 1995). Bechtold et al. (1995) find $40h^{-1} < R < 280h^{-1}$ kpc at $z \sim 1.8$ from Q1343+266A & B separated by 9.5". Dinshaw et al. (1995) derive $270 < R < 860h_{100}^{-1}$ kpc at $z \sim 0.8$ from FOS observation of 0107-025A & B separated by 1.44'. The latter dimensions are much larger than typical galactic sizes and rule out standard CDM minihalo models as well as models of freely expanding or pressure confined clouds.

Association of Ly$\alpha$ systems with galaxies is difficult to investigate directly, except at low redshift, where HST observations are now available; galaxies are detected routinely using multi-object spectroscopy. Lanzetta et al. (1995) have found 11 galaxies coincident in redshift with an absorption system of which nine are Ly$\alpha$-only systems and two have associated CIV absorptions. They report that (i) most of the $z \sim 0.3$ luminous galaxies are surrounded by gaseous envelopes of radius $\sim 160 h_{100}^{-1}$ kpc and (ii) the fraction of absorption clouds lying at a distance smaller than about $150 h_{100}^{-1}$ kpc from a luminous galaxy is at least 0.35$\pm$0.10 and likely 0.65$\pm$0.18. Le Brun et al. (1995) have detected 19 additional galaxies associated with a Ly$\alpha$-only system and derived impact parameters as large as $500 h_{100}^{-1}$ kpc. However anti-correlation between Ly$\alpha$ rest-frame equivalent width and the impact parameter is found only for the stronger lines. Moreover most of the studied Ly$\alpha$ lines have $W_{\rm r} > 0.30$ Å whereas the bulk of the Ly$\alpha$ lines has $W_{\rm r} < 0.3$ Å (Morris et al. 1991). The questions of whether weak absorptions are associated with galaxies or are completely unrelated to them (Morris et al. 1993) and what is the spatial distribution of the Ly$\alpha$ gas are still open.

In this respect, the search for Ly$\alpha$ clouds in underdense regions and in voids of galaxies is of particular interest. At high redshift there is no clear evidence for regions devoid of lines (Carswell & Rees 1987, Duncan et al. 1989) although a few examples are known (Dobrzycki & Bechtold 1991, Rauch et al. 1992). At low redshift, very recent HST observations of bright nearby AGNs by Stocke et al. (1995) have revealed one weak ($W_{\rm r}$ = 36 mÅ) Ly$\alpha$ line with no galaxy within 6 Mpc.

In order to shed light on this issue, we have performed $N$-body simulations of the kinematics of the dark matter in the universe, introducing a description of the warm photoionized gas. This gives a new framework in which observations can be interpreted. In a previous paper (Petitjean et al. 1995) we have shown that Ly$\alpha$ gas traces the dark matter filaments. At low redshift, 25% of the lines should arise in extended envelopes of luminous galaxies of radius of the order of 1.5 Mpc, the remainder arising in gas independent of any galaxy and at a distance of 2 to 7 Mpc from the first neighbour galaxy. Here, we describe in detail our procedures and study the evolution with redshift of the number density of lines. Technical details about the simulations are given in Section 2. Results are given in Section 3 and conclusions are drawn in Section 4.

## 2. Simulations

We have improved and adapted the particle-mesh (PM) code described in detail by Kates et al. (1991) and Klypin & Kates (1991) to include the effects of photoionization. The PM code simulates the kinematics of the dark matter and the thermodynamics of baryons with periodic boundary conditions. The simulations reported here used $128^3$ particles on a $256^3$ grid. The choice of the length scales to be assigned to the simulation box and to the cells has to be considered carefully. At $z = 0$, structures are in the non-linear regime for scales smaller than 10 Mpc (assuming h = 0.5). Unless otherwise stated, all results reported here were obtained using a simulation box of 12.8 Mpc, corresponding to a comoving cell size of $l_c = 50$ kpc. The physical cell size is of course $l_c/(1 + z)$

The mass of one particle, $M_p$, is equal to the mass of the matter distributed in a comoving volume of 8 cells with the mean background density today, $\rho_0$:

$$M_p = \rho_0 (2l_c)^3 \tag{1}$$

If $M_p$ is smaller than the Jeans mass $M_J$, each particle can be considered as a non-collapsing, diffuse gas element, and the concentration of a sufficient large number of such elements describes collapsing regions (filaments, merging processes etc..) Let us estimate the Jeans length for a gas with a baryonic mass density $\rho_{gas}$ and a temperature $T$, expressed in units of $T_4$ ($T = 10^4 T_4$):

$$\lambda_J \approx \left(\frac{5\pi R T \Omega_b}{3 G \rho_{gas} \mu}\right)^{\frac{1}{2}} \approx 1201 \text{kpc} \left(\frac{\rho_0}{\rho_{gas}}\right)^{\frac{1}{2}} \Omega_b^{\frac{1}{2}} \left(\frac{T_4}{\mu_H}\right)^{\frac{1}{2}} \tag{2}$$

where $R$ is the gas constant, $G$ the gravitational constant, $\mu$ the molecular weight per hydrogen atom and $\Omega_b = <\rho_{gas}>_0 /\rho_0$ the average ratio of baryonic to dark-matter density. The requirement $M_p < M_J$ leads to the relation

$$2l_c \leq 1.201 \text{Mpc} \left(\frac{T_4}{\mu_H}\right)^{\frac{1}{2}} (1+z)^{-\frac{1}{2}} \tag{3}$$

This relation is well satisfied up to high redshifts for our cell size.

Our code averages the density field over the cell length. Let us regard one cell as an "elementary cloud." Let $n_1 = \rho/\rho_0$ denote the local density contrast. The maximum density contrast for an elementary cloud to be Jeans stable is then $n_{1,\max} \approx (\lambda_{J,0}/l_c)^2 \approx 150$ where $\lambda_{J,0}$ is the Jeans length for $\rho = \rho_0$ at $z = 0$. This density contrast corresponds to a HI column density $N_{c,\max}$ through the elementary cloud. For a nearly vanishing UV-background flux at $z = 0$ one obtains

$$N_{c,\max} \approx n_{1,\max} n_{H,0} l_c = \frac{\lambda_{J,0}^2}{l_c} n_0 \tag{4}$$

where $n_{H,0}$ denotes the present hydrogen number density.

For $z > 0$ and an UV flux with an amplitude $f_0$ in units of $4\pi J_{21}$ with $J_{21} = 10^{-21}$ erg s$^{-1}$ sr$^{-1}$ cm$^{-2}$ it follows,

$$N_{c,\max} \approx \frac{(n_{H,0})^2}{7.32 f_0} \lambda_{J,0} \left(\frac{\lambda_{J,0}}{2l_c}\right)^3 \tag{5}$$

$$\approx 7.7\, 10^{15} f_0^{-1} \left(\frac{50 \text{kpc}}{l_c}\right)^3 \text{cm}^{-2} \tag{6}$$

Clouds with HI column densities smaller than $10^{16}$ cm$^{-2}$ at $z = 0$ can thus be considered as Jeans stable. Clouds with higher column densities are unstable and should undergo fragmentation. This happens only in the very central part of the dark matter halos. Higher resolution should be used to describe the structure of the densest regions.

According to our procedure to be described below, the minimum gas density for which a column density is computed is the mean baryonic density. This density corresponds to the smallest HI column density that the simulation can reproduce, $N_{c,\min}$. For $z = 0$, $N_{c,\min}$ is smaller than $10^{11}$ cm$^{-2}$. For non-vanishing UV-background flux the $z$-dependence is given by

$$N_{c,\min}(z) \approx \frac{2l_c n_{H,0}^2 (1+z)^5}{7.32 f_0} \tag{7}$$

It follows that for $z > 3.3$ and $f_0 = 0.5$, the smallest HI column density to be found in a simulation with our resolution $l_c = 50$ kpc is $N(HI) \sim 10^{13}$ cm$^{-2}$.

The positions, velocities, temperatures, and gas densities associated with the particles are followed starting from an initial spectrum (details to be discussed below) and utilizing the Zel'dovich approximation as long as it remains valid.

The baryonic material is approximated as following trajectories of the dark matter with *constant* $\Omega_b = 0.1$. This approximation is quite accurate in all regions prior to the first pancakes (or "shocks"). After shocking, the accuracy of the constant-$\Omega_b$ approximation is of course degraded in regions where secondary shocking and/or very efficient cooling take place, but accuracy in these regions is not critical here. The conditions for maintaining constant $\Omega_b$ *are* well satisfied in Ly$\alpha$ clouds: The potential wells of the dark matter structures must be deep enough to prevent the gas from escaping, but still shallow enough to keep the central matter density below levels permitting strong cooling and subsequent star formation to occur.

As perturbations grow, the first objects start to collapse, producing shocks in the gas. Our code seeks shocks in the form of shell crossings. More precisely, a particle is labelled "shocked" if either of two criteria are met: (1) If the Jacobi determinant of the transformation from Lagrange (particle) to Euler (grid) coordinates becomes negative. (2) If the particle enters a shocked region, defined as entering a cell on a coarse grid of cell size $2l_c$ containing at least one shocked particle. In either case, we attempt to define a local velocity field **U** on the coarse grid, and if successful, we assign to the gas associated with the particle a temperature according to $kT = \mu_H m_H (\mathbf{v} - \mathbf{U})^2/3$, where **v** is the particle velocity, $\mu_H$ the molecular weight and $m_H$ the mass of the hydrogen atom. (We take primordial abundances throughout.) It can happen in a small box that the temperature assigned to the particle is below the minimum temperature

(here 5000 K) to be considered "cold." In this case, we return the particle to the "unshocked" pool of particles, where it will most likely again be detected as shocked at some later timestep. The particle is also returned to the unshocked pool in those rare cases in which too few neighboring particles are present to determine **U**. The accuracy of this procedure in comparison to hydrodynamic simulations of pancake collapse is discussed in Kates et al. (1995).

After successful assignment of temperature, we follow the thermal history by integrating the energy equation along particle trajectories in the form

$$\frac{dT}{dt} = (\gamma - 1)\left[\frac{T}{n_H}\frac{dn_H}{dt} - \frac{\mu_M}{\mu_H}\frac{1}{kn_H}(\Lambda_{tot} - \Gamma_{phot})\right] \quad (8)$$

where $\gamma = 5/3$ and $n_H$ is the number density of hydrogen atoms. Here, $\Lambda_{tot}$ is the total cooling rate (Klypin & Kates 1991, Kates et al. 1991) including radiative and dielectronic recombination, bremstrahlung, collisional excitation (Black 1981) and Compton cooling (the latter is the most important process at $z > 5$).

The *heating* rate due to photoionization by the ionizing background (i.e., from QSOs and galaxies) is denoted in Eq.(8) by $\Gamma_{phot}$. It is computed as the sum of the contributions by HI, HeI and HeII assuming ionization equilibrium

$$\xi_i n(i) + \xi_{ei} n(i) n_e = n(i+1) n_e \alpha(i) \quad (9)$$

where $i$ = (HI, HeI, HeII) and $i + 1$ = (HII, HeII, HeIII), $\xi_i$ and $\xi_{ei}$ are the rates of the photo- and collisional ionization respectively, $n_e$ and $n_i$ the electronic and ionic number densities, and $\alpha(i)$ denotes the recombination rate towards $(i)$. The photo-ionization rate for each $(i)$ is given by

$$\xi_i = \int_{\nu_i}^{\infty} \frac{F_\nu \sigma_i(\nu)}{h\nu} d\nu \quad (10)$$

where $h\nu_i$ is the ionization energy (13.6 eV for HI). Opacity effects are taking into account writing

$$F_\nu = F_{0\nu} \exp(-\tau_\nu) \quad (11)$$

with

$$\tau_\nu = \int_x^{x_0}\left[\sum_i n_i(x)\sigma_\nu(i)\right]dx \quad (12)$$

The photoionization cross-section of each of the species, $\sigma$, decreases with $\nu$ faster than $\sigma_\nu \propto \nu^{-3}$. Therefore we approximate the total optical depth $\tau_\nu$ by $\tau_\nu$(HI), $\tau_\nu$(HeI) and $\tau_\nu$(HeII) in the range $\nu_{HI} \leq \nu < \nu_{HeI}$, $\nu_{HeI} \leq \nu < \nu_{HeII}$ and $\nu_{HeII} \leq \nu$ respectively. Therefore,

$$\xi_i(x) = F_0 \exp(-\tau_i(x))G_i^B \quad (13)$$

where

$$G_i^B = \int_{\nu_i}^{\infty} \frac{F_\nu}{F_0}\sigma_i(\nu)d\nu \quad (14)$$

The heating rate is written as

$$\begin{aligned}\Gamma_i &= n_i \int_{\nu_i}^{\infty} \frac{F_\nu(x)h(\nu - \nu_i)}{h\nu}\sigma_i(\nu)d\nu \\ &\approx F_0 \exp(-\tau_i(x))G_i^B \epsilon_i^B n_i \end{aligned} \quad (15)$$

where $G_i^B$ and $\epsilon_i^B$ are respectively the ionization rate and heating paramter defined and given by Black (1981).

To estimate the heating rate, we model each particle in accordance with the PM picture as an independent uniform cloud of volume $(l_c)^3$ and density taken as the local density given by the code. Then we can assume $\tau_i \approx \sigma_i n_i l_c/2$, which corresponds roughly to the optical depth near the center of this cloud.

We solve equations (9) together with mass and charge conservation and derive the heating and cooling rates assuming $n_{He} = 0.1\ n_H$ and $n_e \approx n_{HII}$. The density ratios $n(i+1)/n(i)$ are computed self-consistently, as described above.

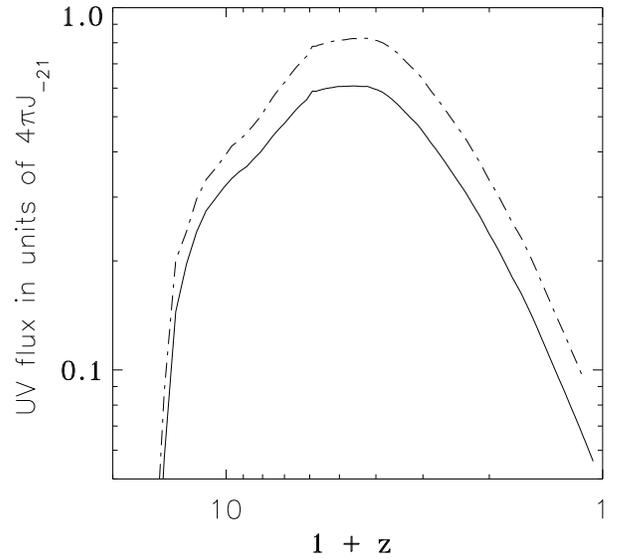

**Fig. 1.** The UV flux generated by the code as function of $1 + z$ with $f_0 = 0.57$ (solid curve) and $f_0 = 0.78$ (dashed curve).

The shape of the ionizing spectrum is modelled as $F_\nu \propto F_0 \nu^{-1}$. $F_0 = f_0 4\pi J_{21}$ is the ionizing flux at 13.6 eV and depends on redshift. It is assumed to be proportional to the rate, $\Delta m(T < 5000K; z)$, at which baryonic material cools below 5000 K in the simulation:

$$\begin{aligned}f_0(z) &= C_{cool}\Delta m(T_4 < 0.5; z) \\ &+ f_0(z + \Delta z)\left(\frac{1+z}{1+z+\Delta z}\right)^4 \end{aligned} \quad (16)$$

where $C_{cool}$ is a factor of proportionality. This simulates an ionizing flux produced as a consequence of gas collapsing in the dense regions. In this case most of the photons just above 1 Ryd would have been produced by massive stars (see also Madau & Shull 1995). The redshift dependence is thus consistently given by the simulation; the only free parameter is the normalization (thus $C_{\rm cool}$) which is done assuming some value of $f_0$ at $z = 5$. The flux as a function of $1 + z$ is given in Fig. 1 for two simulations with $f_0 = 0.57$ (solid curve) and $f_0 = 0.78$ (dashed curve). In both cases, the flux remains nearly constant near its maximum value between $z = 5$ and $z = 2.5$. Below $z = 2.5$ the flux decreases approximately as a power law $\propto (1 + z)^2$. The slope becomes steeper at $z \to 0$.

The origin and intensity of the UV-background has been the subject of intensive investigations (e.g. Bechtold et al. 1987, Miralda-Escudé & Ostriker 1990, Madau 1992). Although uncertainties are large and the interpretation questionable, modelling of the proximity effect gives a rough estimate of the intensity at high redshift (e.g. Bajtlik et al. 1988, Lu et al. 1991, Bechtold 1994) $f_0 \sim 1$ in the range $2 < z < 3$. Somewhat smaller values ($f_0 = 0.5$), although depending on the shape of the ionizing spectrum, have been derived from analysis of the statistical properties of metal line systems (Petitjean et al. 1992). At smaller redshift, only tentative estimates or upper limits are possible using the proximity effect (Kulkarny & Fall 1993, $0.006 < f_0 < 0.04$ at $z \sim 0.5$) or detection of H$\alpha$ emission from nearby clouds (Songaila et al. 1989, Corbelli & Salpeter 1993, Maloney 1993; $0.01 < f_0 < 0.1$ at $z \sim 0$). Our results are consistent with what is expected for the time evolution. The value at low redshift is larger than the upper limit of Kulkarny & Fall (1993) by a factor of two but within the range from other determinations.

## 3. Results

### 3.1. Fluctuation spectrum

The dynamics of large-scale structure formation is highly sensitive to the initial fluctuation spectrum. Confrontation of models with observations (see for example Gottlöber et al. 1994 and Kates et al. 1995) strongly favor scenarios with broken scale invariance ("BSI"), mixed dark matter models, or low-mass models over "standard", COBE-normalized CDM. In particular, CDM seems to imply an excess of virialized, massive ($M > 10^{13} M_\odot$) objects and too large velocity fields at small scales (as measured for example by relative pairwise velocities.) Nevertheless, one crucial property of COBE-normalized CDM convinced us to use it in the present simulations: the biasing factor of approximately unity.

Most of the results reported here depend sensitively on an accurate representation of the statistics of the *baryonic* density field. Now, as in any non-hydrodynamic simulation, the baryonic distribution is to be inferred from the dark-matter distribution. Normally, if one is interested in large-scale structure and its observational consequences, a biasing factor may be introduced by some prescription, such as by searching for high peaks of the dark-matter distribution. This procedure roughly simulates the effects of baryonic condensations, and is thus reasonably well suited to a prediction of galaxy statistics. However, since it averages baryonic concentrations on a galactic scale, it does not appear to be suitable for an estimate of the density in Ly-$\alpha$ clouds.

With a biasing factor of unity, we may safely assume a *constant* ratio of baryonic to dark-matter density without introducing a systematic error in the baryonic density at the scales of interest. Within the limited spectral range resulting from the limited size of the computational box (see §2), the spectrum of most viable models differs from that of CDM by a nearly constant factor. The main features of CDM leading to an excess of massive structures will not be seen. Excessive small-scale power would admittedly be expected to influence our results somewhat, but certainly less than a systematic underestimate of the baryon density in clouds. Moreover, the amplitude of the power spectrum determines in general the epoch of the formation of the first structures and thus the redshift dependence of the UV flux. Smaller amplitude implies later structure formation. The UV peak is lower, but the distribution is shifted to later epochs, which can lead to a relatively *higher* background flux at small redshift.

### 3.2. Probing the dark matter filaments

From Fig. 1 one can see that the UV flux increases rapidly after $z \approx 10$. This rise corresponds to the redshift at which the first structures appear.

Fig. 2 illustrates the evolution of the cloud distribution. With decreasing $z$, a trend toward concentration of the gaseous matter and a decrease in the number density is evident. The structures seen form the usual network of filaments connected by nodes where galaxies and groups of galaxies are to be found. The neutral hydrogen column density is computed through the slice, and contours of constant neutral hydrogen column density $N({\rm HI}) = 10^{13}$ cm$^{-2}$ are shown. It can be seen that the Ly$\alpha$ gas traces the gravitational potential wells.

### 3.3. Evolution of the line number density

We now wish to compute the line number density evolution from our simulated, periodic universe. One copy of this universe consists of a box of comoving size $L$ containing $n_{\rm c}^3$ equally spaced cells and $n_{\rm p}^3$ particles. For this computation, we imagine that a large number of light rays from ficticious, distant QSO's traverse our universe exactly once during a simulation timestep. (This condition fixes the timestep, of course). We therefore have a collection of many ray segments for each timestep. (A large number of rays is required for good statistics; in fact the number of rays exceeds the number of timesteps). Each ray is constructed by randomly choosing one ray segment per timestep from this collection. This procedure is intended to minimize the effects of periodicity and is in fact quite effective, as discussed in the following subsection. Rays intersect the x-y plane at cell vertices.

Now suppose we have all the simulation data in the box at some time $t_0$, corresponding to a redshift $z_0$. We then perform

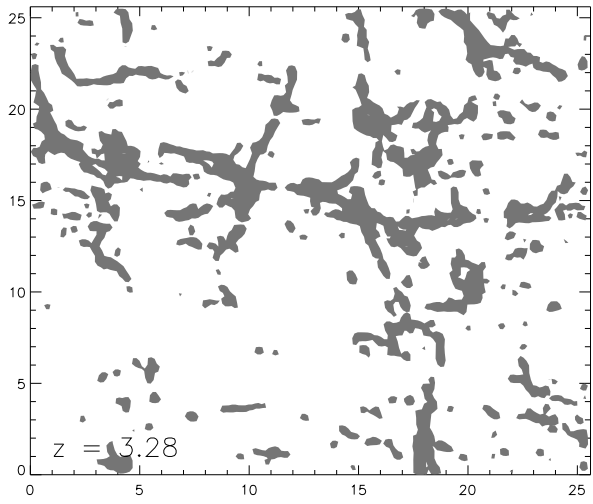

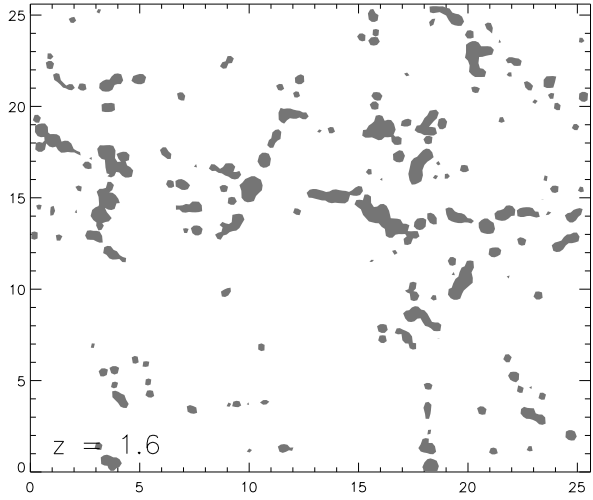

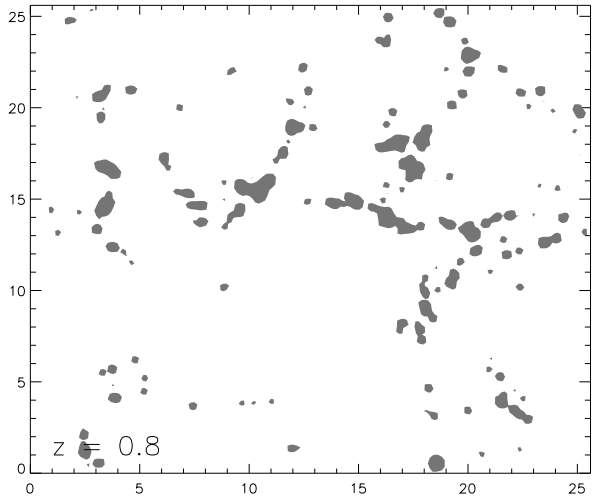

**Fig. 2.** Spacial distribution of Ly$\alpha$ clouds in a 2 Mpc slice of a (25 Mpc)$^3$ simulation box at redshifts $z$ = (3.28, 1.6, 0.8) respectively. Gray contours show regions where $N$(H I) $> 10^{13}$ cm$^{-2}$ through the box.

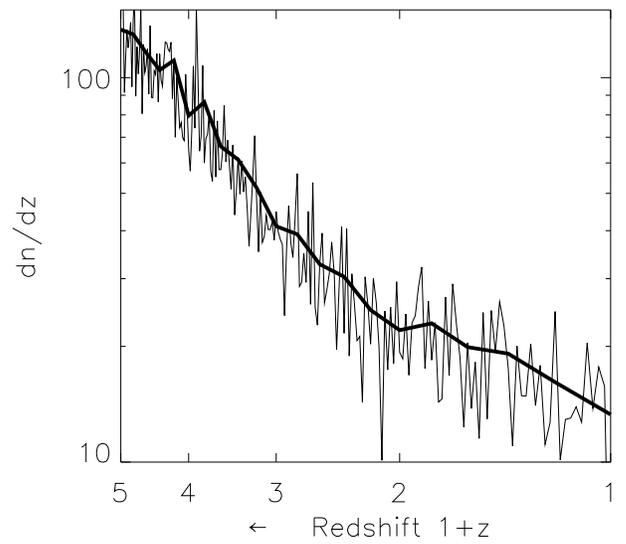

**Fig. 3.** $z$-Dependence of the line number density $dn/dz$ for lines with $N$(HI) $> 10^{14}$ cm$^{-2}$

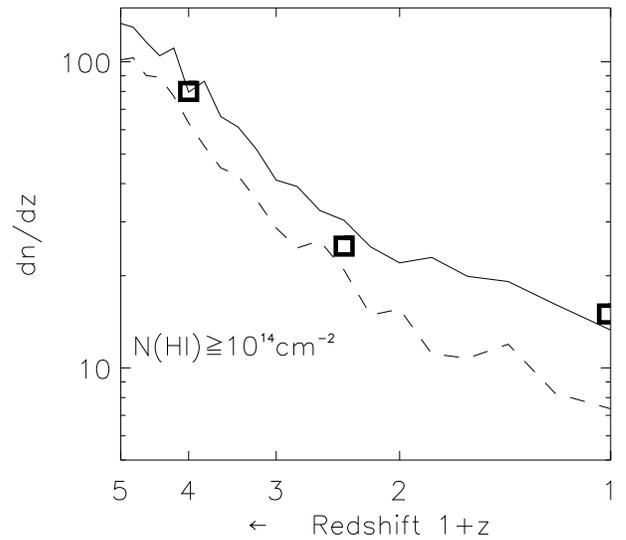

**Fig. 4.** $z$-Dependence of the line number density $dn/dz$ for lines with $N$(HI) $> 10^{14}$ cm$^{-2}$. Solid and dashed lines are for $f_0 = 0.57$ and 0.78 respectively. Observations are overplotted by open squares.

one timestep of the simulation, obtaining the data at time $t_1$ (redshift $z_1$). We may now construct the absorption line spectrum for redshifts between $z_0$ and $z_1$ along the line of sight to our fictitious QSO: As mentioned above, the timestep $\Delta t \equiv t_1 - t_0$ is taken as the time required for a light ray to traverse the comoving length $L$ starting at $z_0$. It turns out that $\Delta t$ is always comfortably less than the maximum allowable timestep for numerical stability, i.e., particles move by less than a cell length within $\Delta t$. However, the time scale for temperature changes according to

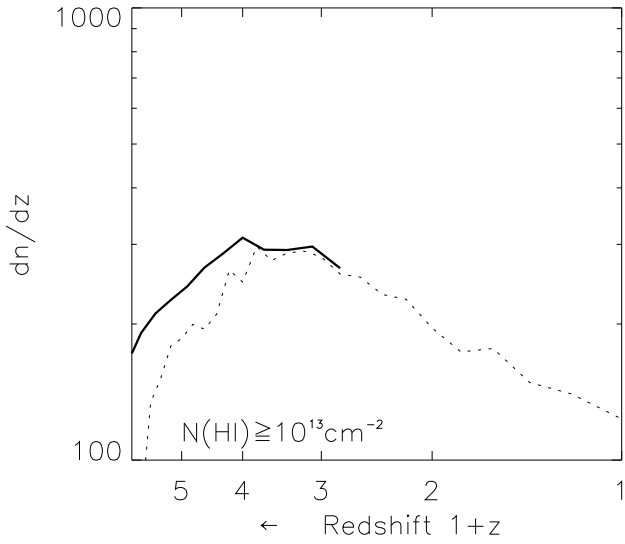

**Fig. 5.** z-Dependence of of the line number density $dn/dz$ for lines with $N(HI) > 10^{13}$ cm$^{-2}$ for two numerical resolutions, $l_c = 50$ kpc (dotted line) and 25 kpc (solid line)

Eq. (8) may be comparable to or shorter than $\Delta t$. We therefore integrate Eq. (8) along the ray with steps $(\Delta z)_T \approx 10^{-3}\Delta z$, where $\Delta z \equiv z_1 - z_0$.

We define *computational cells* of $2^3$ grid cells, i.e., $n_c/2$ per ray. We assign to each computational cell $i$ along the line of sight a redshift $z_i$ between $z_0$ and $z_1$ (corresponding to a time $t_i$) by interpolation as follows: The comoving distance traversed by the light ray between $z_0$ and $z_i$ is simply $i$ times the computational cell size $2l_c$, and $z_i$ may then be obtained from

$$i \cdot 2l_c = \frac{2c}{H_0}\left(\frac{1}{\sqrt{1+z_i}} - \frac{1}{\sqrt{1+z_0}}\right) \quad (17)$$

where $H_0$ denotes the Hubble constant and $c$ the speed of light. Eq. (8) is then integrated for computational cell $i$ up to $t_i$. This procedure is iterated until the entire ray segment is covered by computational cells. With each computational cell, we associate the precise redshift $z_i$ and all of the particle data (position, temperature, density) necessary for obtaining that computational cell's neutral hydrogen content and thus its contribution to the Ly$\alpha$ line number density. Finally, we compile all the ray data for each ficticious QSO and store it for later analysis.

We note for further reference that the resolution $\delta z$ of line contributions is clearly limited by the computational cell length: $\delta z \approx 2H_0 l_c (1+z)^{3/2}/c$.

The z-dependence of $dn/dz$ is determined by averaging over our large sample of rays. Computation of $dn/dz$ on any particular ray at some given redshift $z = z_i$ requires averaging over the contributions of all particle-clouds to the mean column density in computational cell $i$. We imagine that computational cell $i$ contains many randomly spaced clouds with different temperatures and densities and that the distribution of sizes is restricted by the condition of pressure confinement. The averaging is thus to be weighted by the probability that the ray passes through a cloud. The mean column density $N_c(z)$ in one computational cell at $z$ is

$$N_c(z) = \frac{\sum N_i(z) w_i}{\sum w_i} \quad (18)$$

where the $N_i(z)$ denote the column density of each particular cloud in a computational cell at redshift position $z$. We recall that using a PM code each particle mass is distributed over a certain volume surrounding the particle situated in the center of this volume. In the case considered here this volume is equal to the volume of one grid cell (not computational cell). The overlapping parts of these volumes define the mass contributions of nearby particles to the local density of the centered particle. Therefore, the "column density" of each cloud is formally equal to the density times the cell size. The $w_i$ in (18) are the probabilities that the ray crosses the cloud $i$. These probabilities are proportional to the cross section of the clouds, i.e. $\propto r_i^2$ where the $r_i$ are the radii of the clouds. The number density in a cloud is given as

$$n_i = \frac{3m_i}{4\pi m_p r_i^3} \quad (19)$$

where $m_H$ denotes the mass of a hydrogen atom and $m_i = m_0$ is the known mass of a cloud in the given simulation. Constant pressure leads using $p = k_B \mu T_i n_i = const.$ to the relation $r_i = \kappa T_i^{1/3}$, where $\kappa$ is a constant, and therefore to

$$N_c(z) = \frac{\sum N_i(z) T_i^{\frac{2}{3}}}{\sum T_i^{\frac{2}{3}}} \quad (20)$$

which is the mean column density contribution from the the computational cell under consideration on the simulated ray.

We may define an effective cross section $\pi r_{\text{eff}}^2$ of the computational cell as the average cross section of the particle clouds contained in the computational cell weighted by their column densities:

$$r_{\text{eff}}^2 = \kappa^2 \frac{\sum N_i T_i^{\frac{2}{3}}}{\sum N_i} \quad (21)$$

The total volume $(4\pi/3)R_{cc}^3$ of all particle clouds in the computational cell is $\sum r_i^3 = \kappa^3 \sum T_i$. The corresponding cross section is $\pi R_{cc}^2 = \kappa^2 \pi (\sum T_i)^{2/3}$.

Therefore, the probability to find a line with column density $N_c(z)$ in the computational cell at redshift $z$ on ray $j$ is

$$W_j(z) = \left[\frac{\sum_i N_i T_i^{\frac{2}{3}}}{\sum_i N_i} \frac{1}{(\sum_i T_i)^{\frac{2}{3}}}\right]_{j,z} \quad (22)$$

Finally, we estimate the line density $dn/dz(z)$ for any given threshold column density $N_{HI}$ by summing over all rays $j$ those lines at redshift $z$ whose column densities $N_c(z)$ exceed $N_{HI}$, weighted by the probability $W_j(z)$.

Fig. 3 presents the results obtained for the z-dependence of the number density of clouds per unit redshift for clouds with $N_{HI} \geq 10^{14}$ cm$^{-2}$. The thick line presents the same results obtained by smoothing over redshift intervals of $(\Delta z)_{\text{smoothing}} = 0.2$. For that line the following fit formulae can be given

$$\frac{dn}{dz} = 16.8(1+z)^{0.8} + 0.687(1+z)^{2.5} \qquad (23)$$

In Fig. 4 the line number density per unit redshift interval for column densities exceeding $N_{HI} = 10^{14}$ cm$^{-2}$ is plotted for the two different flux amplitudes used in Fig. 1 and compared with the data (square symbols). The latter are taken from compilations by Lu et al. (1991) and Petitjean et al. (1993) at high redshift and from HST observations by Bahcall et al. (1993). The computed number is also consistent with observation at $z > 4$ (Williger et al. 1994) although the latter are still scarce. The upper (solid) curve in Fig. 4 corresponds to the lower (solid) curve in Fig. 1. Obviously the upper curve, corresponding to $f_0 = 0.57$ yields a better fit to the observations.

In Fig. 5, the line number density per unit redshift interval for column densities exceeding $N_{HI} = 10^{13}$ cm$^{-2}$ is plotted for the flux amplitude $f_0 = .57$. The thin solid line corresponds to $l_c = 50$ kpc. In the range from $z = 0$ to $z = 2.5$, $dn/dz$ behaves roughly like $(1+z)^{0.7}$. The apparent drop for $z > 2.5$ illustrates the redshift dependence of the resolution implied by Eq. (7). In order to extend the redshift range for $N_{HI} = 10^{13}$ cm$^{-2}$, we carried out simulations with $l_c = 25$ kpc. The result for corresponding $f_0$ and the same $N_{HI}$ limit is shown as the thick solid line in Fig.5. To perform this simulation with higher resolution we had to decrease the box size. The scale comparable to that box size becomes non-linear already at redshift $z \approx 1$. For this reason, the range of redshift shown is limited to $z > z_{nl} > 1.5$. The observed number of such lines is not well known yet due to the difficulties in observing weak lines at high enough resolution. However very recent observation with the Keck telescope show that at $z \sim 3$ the number density of lines with $N > 10^{13}$ cm$^{-2}$ is about 350 per unit redshift (Songaila et al. 1995). At low redshift the statistics of weak lines is very poor. From the Morris et al. (1991) and Bahcall et al. (1991) data however it is possible to infer that the number per unit redshift of lines with $W_r > 25$ mÅ is larger than 100 at $z < 0.15$. Our computed numbers for $f_0 = 0.57$ are a factor of two larger. However this can be fine-tuned using a somewhat larger $f_0$ value. Given the limitations inherent to the simulation (see above), we consider that the discrepancy is not significant. It is very important to note that the bulk of the Ly$\alpha$ forest has $N(HI) < 10^{14}$ cm$^{-2}$ and that those lines evolve less rapidly than the stronger lines. The latter effect may be present in the data (Bechtold 1994, Bahcall et al. 1995).

Fig 6 presents the column density distributions for a variety of cell lengths, redshifts, and fluxes: In general the curves reproduce well the observed slope of $\beta \approx 1.7-1.8$ for $z \approx 3$. In both curves representing $z = 3$ with low flux, a clear flattening is evident toward lower column densities. The flattening begins at higher $N$ in the $l_c = 50$ kpc simulation than in the $l_c = 25$ kpc simulation, consistent with the above remarks concerning resolution. Higher flux (for $l_c = 25$ kpc) decreases the flattening at low $N$, in agreement with the prediction of Eq. (7). Further implications will be discussed in the Conclusions.

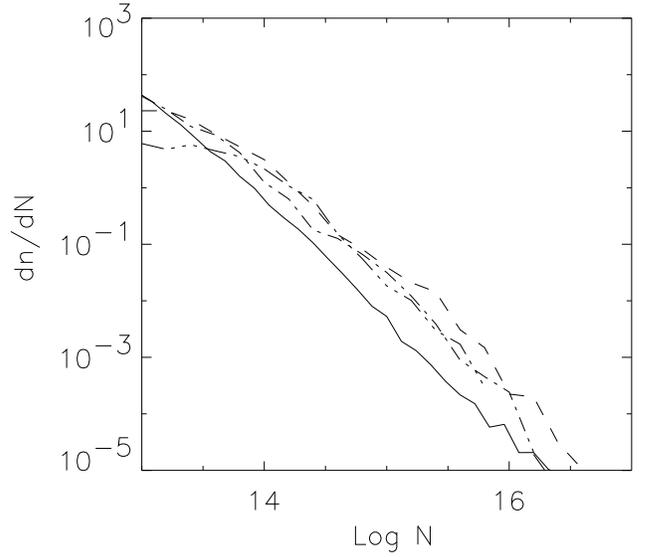

**Fig. 6.** Column density distribution $dn/dN$ for assorted cell length, redshift, and flux values. solid line: $l_c = 50$ kpc, z=0.5, $f_0 = 0.57$; dashed-multidotted line: $l_c = 50$ kpc, $z = 3$, $f_0 = 0.57$; dashed line: $l_c = 25$ kpc, $z = 3$, $f_0 = 0.57$; dashed-dotted line: $l_c = 25$ kpc, $z = 3, f_0 = 1.1$

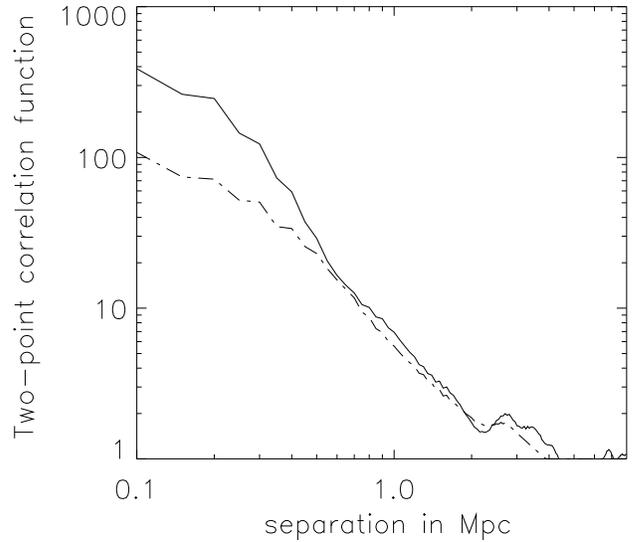

**Fig. 7.** Two-point correlation function $\xi_c(x)$ for clouds with $N_c = 10^{14} cm^{-2}$ (solid line) and for clouds with $N_c = 10^{13} cm^{-2}$ (dashed line) at z=0.

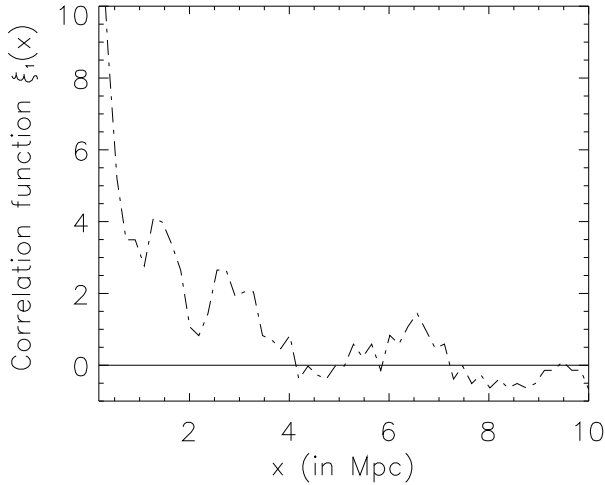

**Fig. 8.** One dimensional correlation function $\xi_1(x)$ for the absorption lines with $N_{HI} > 10^{14} cm^{-2}$ along the l.o.s. between z=3 and z=3

### 3.4. Two point correlation

Our computation is obviously insensitive to correlations on a spatial scale comparable to the box size; indeed, our ray segment procedure is designed to eliminate any spurious correlations resulting from the periodic box. Nevertheless, we may hope to detect correlations at some fraction of the box size:

Fig. 7 shows the two-point correlation functions for clouds with $N_{\rm HI} = 10^{14}$ cm$^{-2}$ (solid line) and for clouds with $N_{\rm HI} = 10^{13}$ cm$^{-2}$ (dashed line) at $z = 0$. In both cases the correlation functions are bi-modal. A component with slope $> -1$ is extended till separations of about 4 Mpc and indicates the contribution to filamentary and sheet-like structures. In the upper correlation function a spherical component is well expressed (slope $< -2$). The signal at small separations for the correlation function related to $N_{\rm HI} = 10^{14}$ cm$^{-2}$ is much stronger than for the distribution of low density clouds. Both functions show a component with constant slope $\approx -1.75$ which is comparable with the slope of the two-point correlation function for the galaxy distribution. That might be understood as an indication of the association of those clouds with galaxies.

In Fig. 8, the one-dimensional correlation function for the line distribution along a ray between $z = 3$ and $z = 2$ is shown for absorption lines with column densities exceeding $10^{14}$ cm$^{-2}$. The clear correlation out to a radius of about 4 Mpc is in good accordance with the two-point correlation function discussed above and with recent results of (Christiani et al. 1995). The signal becomes weaker if we include the contribution of low-density lines.

## 4. Conclusions

Our basic assumption was that the Ly-$\alpha$ clouds could trace the shallower potential wells given by the dark matter distribution. That concerns especially those dark matter structures as filaments and sheets. The quite good agreement of the obtained results with the observational data seems to indicate that the basis of our considerations is quite reasonable.

On the other side this opens the exciting prospective to use Ly$\alpha$ absorption line observations to map the universe in three dimensions and get information about the distribution of dark matter structures and their cosmological evolution.

The weak dependence of the results with respect to the box size (provided it satisfies the initially posed restrictions) leads to the conclusion that the evolutionary effects in the number density of clouds are mainly due to the structure evolution on the smaller scales determining also the flux dependence on the redshift.

Possible correlations of the clouds over scales larger than the box size cannot be investigated in the frame of our considerations due to the limited box size. However, the obtained sufficiently high signal in the one-dimensional correlation function for scales up to $\approx 5$ Mpc has been proved with help of simulations with boxsize of about 25 Mpc. This result deserves further investigation and will be considered in a forthcoming paper.

Incidentally, comparison of Ly-$\alpha$ data with sufficiently realistic (i.e., hydrodynamic) simulations could be used in principle to "normalize" the power spectrum within the dynamical range. However, if one characterizes the "normalization" by an overall amplitude, this amplitude will be overestimated in a small box due to truncation of long-wavelength modes. The relation between box size and time scales for structure formation (at constant amplitude) has been studied by Doroshkevich et al. (1995).

Returning to the column density distribution, we saw that smaller cell length $l_c$ and higher flux $f_0 \approx 1.0$ reduce the flattening at low column densities. However, the data for $N_{\rm HI} > 10^{14}$ cm$^{-2}$ seem to favor a smaller flux $f_0 \approx 0.6$. In our simulations, the shielding of UV radiation in overdense regions is only taken into account within a cell. Our value of $f_0$ thus represents an underestimate of the flux that would be present in average or below average density regions (e.g., voids). A more accurate estimate of the behavior of the line distribution at low column densities would presumably need to take into account the distribution of small structures forming within "unshocked" regions, which were not included in our simulation. The contribution of such structures within voids could be estimated using a "mini-halo" model in combination with Press-Schechter theory (Mo, Miralda-Escudé, & Rees, 1993).

# References

Bahcall J.N., Bergeron J., Boksenberg A., et al., 1993, ApJS 87, 1
Bahcall J.N., Bergeron J., Boksenberg A., et al., 1995, ApJ (preprint)
Bahcall J.N., Januzzi B.T., Schneider D.P., et al., 1991, ApJL 377, 5
Bajtlik S., 1995. In Bergeron J., Meylan G., Wampler J. (eds) Proc. ESO Workshop, QSO Absorption Lines. Springer, Heidelberg (in press)
Bajtlik S., Duncan R.C., Ostriker J.P., 1988, ApJ 327, 570
Barcons X., Webb J.K., 1991, MNRAS 253, 207
Bechtold J., 1994, ApJS 91, 1
Bechtold J., Crotts A.P.S., Duncan C., Fang Y., 1995, ApJ (in press)
Bechtold J., Weymann E.J., Lin Z., Malkan M.A., 1987, ApJ 315, 180
Black J.H., 1981, MNRAS 197, 553
Bond J.R., Szalay A.S., Silk J., 1988, ApJ 324, 627
Carswell R.F., Rees M., 1987, MNRAS 224, 13P
Cen R., Miralda-Escudé J., Ostriker J.P., Rauch M., 1994, ApJL 437, L9
Chaffee F.H., Foltz C.B., Bechtold J., Weymann R.J., 1986, ApJ 301, 116
Chernomordik V.V., 1995, ApJ 440, 431
Corbelli E., Salpeter E.E., 1993, ApJ 419, 94
Cowie L.L., Songaila A., Kim T.S., Hu E.M., 1995, AJ 109, 1522
Cristiani S., D'Odorico S., Fontana A., Giallongo E., Savaglio S., 1995. In Bergeron J., Meylan G., Wampler J. (eds) Proc. ESO Workshop, QSO Absorption Lines. Springer, Heidelberg (in press)
Crotts A.P.S., 1989, ApJ 336, 550
Dinshaw N., Foltz C.B., Impey C.D., et al., 1995, Nature (in press)
Dobrzycki A., Bechtold J., 1991, ApJL 377, L69
Doroshkevich A.G., Fong R., Gottlöber S., Mücket J.P., Müller V., 1995, MNRAS (submitted)
Duncan R.C., Ostriker J.P., Bajtlik S., 1989, ApJ 345, 39
Duncan R.C., Vishniac E.T., Ostriker J.P., 1991, ApJL 368, 1
Foltz C.B., Weymann R.J., Röser H.J., Chaffee F.J.Jr., 1984, ApJL 281, L1
Gottlöber S., Mücket J.P., Starobinsky A.A., 1994, ApJ 434, 417
Ikeuchi S., 1986, Astrop. & Spa. Sci. 118, 509
Ikeuchi S., Ostriker J.P., 1986, ApJ 301, 522
Kates R.E., Kotok E.V., Klypin A.A., 1991, A&A 243, 295
Kates R.E., Müller V., Gottlöber S., Mücket J.P., Retzlaff, (1995), MNRAS (in press)
Klypin A.A., Kates R.E., 1991, MNRAS 251, 41p
Kulkarny V.P., Fall S.M., 1993, ApJ 413, L63
Lake G., 1988, ApJ 327, 99
Lanzetta K.M., Bowen D.V., Tytler D., Webb J.K., 1995, ApJ 442, 538
Le Brun V., Bergeron J., Boissé P., 1995, A&A (in press)
Lu L., 1991, ApJ 379, 99
Lu L., Wolfe A.M., Turnshek D.A., 1991, ApJ 367, 19
Madau P. 1992, ApJ 389, L1
Madau P., Shull J.M., 1995, preprint
Maloney P., 1993, ApJ 414, 41
Miralda-Escudé J., Ostriker J.P., 1990, ApJ 350, 1
Morris S.L., Weymann R.J., Dressler A., et al. 1993, ApJ
Morris S.L., Weymann R.J., Savage B.D., Gilliland R.L., 1991, ApJL 377, 21
Morris S.L., van den Bergh S., 1994, ApJ 427, 696
Ostriker J.P., Bajtlik S., Duncan R.C., 1988, ApJL 327, L35
Petitjean P., 1995a. In Walsh J.R., Danziger I.J (eds) Proc. ESO Workshop, Science with the VLT. Springer, Heidelberg, p.
Petitjean P., 1995b. In Bergeron J., Meylan G., Wampler J. (eds) Proc. ESO Workshop, QSO Absorption Lines. Springer, Heidelberg (in press)
Petitjean P., Bergeron J., Puget J.L., 1992, A&A 265, 375
Petitjean P., Mücket J., Kates R.E., 1995, A&A 295, L9
Petitjean P., Webb J.K., Rauch M., et al., 1993, MNRAS **262**, 499
Rauch M., Carswell R.F., Chaffee F.H., et al., 1992, ApJ 390, 387
Rees M., 1986, MNRAS 218, 25P
Smette A., Surdej J., Shaver P.A., et al., 1992, ApJ 389, 39
Smette A., et al., 1995, A&A (in press)
Songaila A., Bryant W., Cowie L.L., 1989, ApJ 345, L71
Songaila A., Hu E.M., Cowie L.L., 1995, Nature 375, 124
Stocke J.T., Shull J.M., Penton S. et al., 1995, ApJ (in press)
Tytler D., 1995. In Bergeron J., Meylan G., Wampler J. (eds) Proc. ESO Workshop, QSO Absorption Lines. Springer, Heidelberg (in press)
Tytler D., Fan X., 1994, ApJL 424, L87
Webb J.K., 1987, In Hewitt A., Burbidge G., Fang L.Z. (eds) Proc. Observational Cosmology, IAU Symp. N.124, Kluwer, Dordrecht, p. 803
Williger G.M., Baldwin J.A., Carswell R.F., et al., 1994, ApJ 428, 574
Zel'dovich Y.B., 1970, A&A 5, 84